\documentclass[sigconf]{acmart}

\AtBeginDocument{%
  \providecommand\BibTeX{{%
    \normalfont B\kern-0.5em{\scshape i\kern-0.25em b}\kern-0.8em\TeX}}}

\usepackage{multirow}
\usepackage{makecell}
\usepackage{csquotes}

\copyrightyear{2024}
\acmYear{2024}
\setcopyright{rightsretained}
\acmConference[CHI EA '24]{Extended Abstracts of the CHI Conference on Human Factors in Computing Systems}{May 11--16, 2024}{Honolulu, HI, USA}
\acmBooktitle{Extended Abstracts of the CHI Conference on Human Factors in Computing Systems (CHI EA '24), May 11--16, 2024, Honolulu, HI, USA}
\acmDOI{10.1145/3613905.3644065}
\acmISBN{979-8-4007-0331-7/24/05}

\begin{document}

\title{AI as a Child of Mother Earth: Regrounding Human-AI Interaction in Ecological Thinking} 


\author{Chunchen Xu}
\email{cxu66@stanford.edu}
\orcid{0000-0002-3504-0454}
\affiliation{%
  \institution{Stanford University}
  \streetaddress{}
  \city{Stanford}
  \state{California}
  \country{USA}
  \postcode{94305}
}

\author{Xiao Ge}
\orcid{}
\email{xiaog@stanford.edu}
\affiliation{%
  \institution{Stanford University}
  \streetaddress{}
  \city{Stanford}
  \state{California}
  \country{USA}
  \postcode{94305}
}

\renewcommand{\shortauthors}{Xu \& Ge}

\begin{abstract}

The anthropocentric cultural idea that humans are active agents exerting control over their environments has been largely normalized and inscribed in practices, policies, and products of contemporary industrialized societies. This view underlies a human-ecology relationship based on resource and knowledge extraction. To create a more sustainable and equitable future, it is essential to consider alternative cultural ideas rooted in ecological thinking. This perspective underscores the interconnectedness between humans and more-than-human worlds. We propose a path to reshape the human-ecology relationship by advocating for alternative human-AI interactions. In this paper, we undertake a critical comparison between anthropocentrism and ecological thinking, using storytelling to illustrate various human-AI interactions that embody ecological thinking. We also delineate a set of design principles aimed at guiding AI developments toward fostering a more caring human-ecology relationship. 



\end{abstract}

\ccsdesc[500]{Human-centered computing~HCI theory, concepts and models}


\keywords{Human-ecology relationship; More-than-human; AI; Sustainability; Anthropocentrism; Ecological thinking; Environmental justice; Culture; Storytelling; Design}



\maketitle

\section{Introduction}

\begin{displayquote}{\emph{Every grass and tree has its own language, and the dangling rocks and trees everywhere talk to each other, and at night it is like a strange house of fire burning, and at daytime it is like the sound of the wings of a swarm of insects, and everywhere is lively. 
\null\hfill— Kojiki}}
\end{displayquote} 

The above is a lively non-human world depicted by a Japanese chronicle of myths compiled in the Nara period in the 8th century. Viewing the non-human world as full of life is common in pre-industrial societies and in many historically marginalized groups \cite{watts2013indigenous,mauss2000gift,strathern1988gender}. Accordingly, humans are not the center of the world, but are part of an interconnected web of living beings and matter. Some scholars in humanities and social sciences have used the term ``ecological thinking'' \cite{morton2010ecological,Chisholm2011,code2006ecological} to encapsulate such ideas.


Ecological thinking contrasts with \textit{anthropocentrism}, which represents a distinct set of epistemologies, values, norms, and practices that elevate humans as active agents separate from and exerting control over their socio-physical environments \cite{hoffman2005getting,purser1995limits,thompson1994ecocentric}.\footnote{This definition of anthropocentrism may be perceived as portraying ``human'' as a dubious ``species category'' \cite{malm2014geology}, which carries the risk of essentializing the human identity. The term ``anthropocentrism'' overlooks crucial cultural variations in how people define what it means to be ``human'' \cite{markus2010cultures}. Notably, people in many interdependent cultural contexts do not perceive themselves as separate from other people or their environments \cite{markus2014culture}. We adopt a social constructionist lens \cite{gergen1992social,butler2002gender} and suggest that there is no homogeneous, universal, or static category of humanity. In this paper, we critically refer to ``anthropocentrism'' as an umbrella term encompassing an interrelated cluster of normative assumptions, beliefs (e.g., dichotomies between humans and nature, disembodied way of knowing), and narratives (e.g., mastery and control over nature). We urge readers to consider the limitations and discursive particularities of the term.} In this paper, we argue that anthropocentric thinking is associated with a widely held expectation that AI should be a tool or smart assistant in service of fulfilling individual goals and desires.\footnote{The meanings of ``tool'' vary across cultures, and people's mindsets and meaning-making are integral to understanding AI's impacts. In contexts where ecological thinking is more prevalent, people also construe technologies as tools. Nonetheless, they are less likely to see these tools as separate from themselves or serving their individual goals. Even with existing designs of AI tools or smart assistants, people influenced by ecological thinking may perceive AI differently from those influenced by anthropocentrism.} Instead, we envision AI as \emph{a child of Mother Earth} from the perspective of ecological thinking. We explore how an alternative mode of human-AI interaction can potentially nurture a more caring human-ecology relationship.






\subsection{Cultural Underpinnings of the Human-Ecology Relationship}

In many contemporary industrialized societies, it can be normative to construe natural environments as objective, inert, and fading to the background while moving humans in the center to exert control \cite{mcshane2007anthropocentrism,knappett2008material}. For example, Purdy \cite{purdy2015after} points out that ``a frontier vision of settlement and development, a wilderness-seeking Romanticism, a utilitarian attitude that tries to manage nature for human benefit'' have been implicit assumptions underlying environmental policies in the U.S. Anthropocentric ideas are rooted in historically derived, hierarchy-embedding dualisms such as mind over matter and civilization above wilderness, which have contributed to objectifying nature and using it for human benefit \cite{haila2000beyond,merchant1980death,braidotti2019theoretical}. 


As philosopher Arne Næss argues, a reflective reevaluation of industrialized societies' cultural values and beliefs is imperative for fostering deeper environmental engagement \cite{naess1973shallow}. Contrary to the anthropocentric outlook, humanity ``stands in a constitutive relation to its non-human others'' \cite{marder2013plant}. Amid profound environmental predicaments, recent years have witnessed tremendous interest in ``the de-centering of the metaphysical image of the human'' \cite{marder2013plant}. Many scholars and scientists have explored a potential ontological shift towards more ecological thinking, arguing for profound cultural changes to underpin science, organizational practices, and environmental policies and ethics, as well as technological developments \cite{barad2007meeting,medin2017systems,braidotti2019theoretical,stengers2018another,purser1995limits,BRAGG199693,Sessions1987,haraway2016staying,de2017matters}. Intellectual movements and concepts that are formative to this ongoing change include \emph{feminist environmental scholarship} \cite{merchant1980death,haraway2013cyborg,alaimo2012sustainable,shiva2016staying,plumwood2002feminism,warren1990power}, \emph{deep ecology} \cite{Sessions1987,naess1984defence,seed1988thinking}, \emph{plant-thinking and vegetal being} \cite{marder2013plant,marder2016through,hall2011plants,simard2021finding}, \emph{posthumanism} \cite{wolfe2010posthumanism,braidotti2019posthuman,colebrook2014death}, \emph{new animism} \cite{harvey2005animism,abram2012spell,kohn2013forests}, \emph{relational and process-based ontology} \cite{wildman2010introduction,deleuze1994difference,whitehead2010process}, and \emph{material agency} \cite{knappett2008material,latour2007reassembling}, among other streams of thought and theories across disciplines. We draw on these various bodies of work while acknowledging that they inform, diverge from, and sometimes contest each other in important ways. In particular, our conception of the human-ecology relationship based on care and connection finds its roots in feminist environmental thought, aesthetics and ethics \cite{merchant1981earthcare,griffin1978women,de2017matters,alaimo2016exposed,macgregor2011beyond,barad2007meeting,hooks2000feminist}.


Highlighting humans' interdependence with the non-human world is simultaneously a call to advance social equity by representing epistemologies and practices of diverse groups who tend to skillfully participate in natural environments with care \cite{merchant1981earthcare,selin2013nature,kimmerer2013braiding,bang2018if,cajete1994look}. For example, Sámi people, a nomadic group in northern Scandinavia, create a \textit{cognitive map} to represent their interconnected landscapes \cite{cogos2017sami,Kotva2021}. Traditional contemplation practices help cultivate plant-like attentiveness and receptivity to natural environments \cite{batchelor1994buddhism,kotva2022effort}. Contemporary Japanese Butoh dancers strive to ``erase the body’s outlines'' and resemble wind and clay in rice fields \cite{candelario2019dancing}. Oral languages coupled with specific namings of objects can powerfully enable people to \emph{speak to (rather than about)} the world \cite{abram2012spell}. Ecological thinking is also elegantly captured by the Zen Buddhist saying that ``(the) whole world is a single flower'' \cite{Sahn1992}: Each component of a healthy ecological system is complexly intertwined with each other and each part contains the whole. As such, our work is also grounded in the decolonial and environmental justice literature \cite{mignolo2018decoloniality,kimmerer2013braiding,mcgregor2020indigenous}. 





\subsection{Human-AI Interaction is Intertwined with the Human-Ecology Relationship}
At the nexus of a much-needed cultural shift for sustainability is humans' relationship with AI-based smart technologies. Throughout history, many transformative technologies (such as electricity and the internet) have played a pivotal role in institutionalizing anthropocentrism \cite{elam2022signs,zuboff2023age}, separating engineered environments from nature and bestowing authority upon a ``disembodied form of technological knowing'' \cite{purser1995limits}. Today's smart technologies have similarly inherited manifold anthropocentric ideas \cite{elam2023poetry,elam2022signs}, including an enduring root metaphor of ``the machine'' \cite{merchant1981earthcare} that functions to bolster humans' distinction from and active control of their environments \cite{xiao2024}. The mantra for smart technological developments in countries such as the U.S. has typically centered on assisting individuals in pursuing goals and realizing their potential or helping organizations to be more efficient and competitive. 

For instance, the growing business of smart home technologies builds on assumptions about users' desire for having control over their personal environments, reflecting an undiversified, anthropocentric view on people's relationships with their environments \cite{aarts2009new,remagnino2004ambient}. Likewise, algorithms are designed to purportedly offer a variety of choices and cater to personal preferences on social media or online shopping sites \cite{uchyigit2008personalization}.\footnote{Research suggests that algorithms often result in constraining choices and freedom \cite{talaifar2023freedom,bucher2018if}.} Organizations are incorporating AI into their decision processes to reduce cost and more effectively manage the workforce \cite{newlands2021algorithmic,kellogg2020algorithms}. Consequently, the engineered environments are \emph{miniature daily theaters} where people and institutions rehearse and reinforce anthropocentric ideas that are also fluidly applied to constructing human-ecology relationships writ large \cite{markus2010cultures}. Nature is often regarded as a remote warehouse, an abstract external stakeholder, or a romanticized backdrop for people to record their happiest moments \cite{chang2020social}.

Anthropocentric assumptions are evident even at the intersection of AI and environmental sustainability. Many AI-based products and systems are developed to assist humans in monitoring the environment and efficiently managing resources \cite{reichstein2019deep}. Despite ample enthusiasm about AI's potential for addressing environmental issues, a paucity of work has adopted an ecological thinking perspective. Rarely do developers ask questions such as, ``\emph{How might different animals connect with AI}?'',\footnote{This question was inspired by the book \emph{What Would Animals Say If We Asked the Right Questions?} written by Vinciane Despret \cite{despret2016would}.} ``\emph{What is AI's role in the ecosystem}?'', and ``\emph{Does AI strengthen people's connection with the planet}?''. Deploying AI as tools to solve problems under anthropocentric framings can perpetuate systems of domination over nature. The urgency of change is also prompted by recent criticisms about reverberating environmental consequences due to the large amount of energy and materials needed to sustain rapid AI developments \cite{Itoi2022}. As such, extant approaches might entrench anthropocentrism to the detriment of actively preserving the welfare of the more-than-human world \cite{abram2012spell,de2017matters}. 

It is timely and imperative to reject an illusory demarcation between human society and nature and acknowledge that we are always in nature \cite{Kotva2021,haraway2013cyborg}, and technological systems constantly participate in the ``the council of all beings'' \cite{seed1988thinking}. In recent years, scholars in HCI have started to promote discussions on more-than-human (centered) epistemologies and approaches to AI design \cite{Nicenboim2020,Akama2020,Coskun2022,yoo2023morethanhuman,Jaaskelainen2022,Karana_McQuillan_Rognoli_Giaccardi_2023}. We join these emerging efforts and explore how technologies can be coextensive with the broader natural environment. This pursuit can radically broaden HCI research by ``thinking big'' \cite{morton2010ecological} and revealing humans' and AI's vibrant and precarious interdependence with the ecosystem. If we see ``ecology'' as ``the science of the (Earth's) household''\footnote{The Greek origin of the word ``ecology'' means ``household'' \cite{schwarz2011etymology}.} \cite{merchant1981earthcare} rather than the management of humans' warehouse or playground, then ecological thinking raises an essential question of how humans should participate in the household and relate to each other and to various forms of existence on Earth. 






\section{Reshaping Culture through Redesigning AI}


We first propose a conceptual map to better understand anthropocentrism and ecological thinking.\footnote{Anthropocentrism and ecological thinking encompass multiple ideas rather than being singular concepts. The two dimensions we propose here offer one approach to understanding their systematic differences. However, we may not capture the complete range and variety of conceptual and practical divergences between anthropocentrism and ecological thinking. For instance, our conceptual map implies, but does not thoroughly explore, the notion of equality or inequality among all Earth beings, a crucial distinction between anthropocentrism and ecological thinking. Readers are encouraged to view the conceptual map as an initial exploration rather than a definitive conclusion.} We then introduce the concept of \emph{cultural symbiosis} and propose ways in which AI developments can help restructure humanity's troubled environmental relations toward care and rejuvenation. 



\subsection{Dimensions of Anthropocentrism and Ecological Thinking}



Imagine the differences—in values, experiences, and outcomes—between Jakua, a nature lover who understands the benefits of spending time in nature and who sees nature as providing a stage for outdoor adventures, and Gilaru, another nature lover and adventurer who feels connected to and befriends spiders and frogs and who appreciates stones and waves and everything in between (adapted from \cite{hildaseries}). Jakua's perspective is common in the U.S. population and represents anthropocentric cultural ideals. Gilaru’s approach, in contrast, is based on ecological thinking, which emphasizes interconnectedness between humans and lively more-than-human worlds.

\begin{figure}[!htbp]
  \centering
\includegraphics[width=1.0\columnwidth]{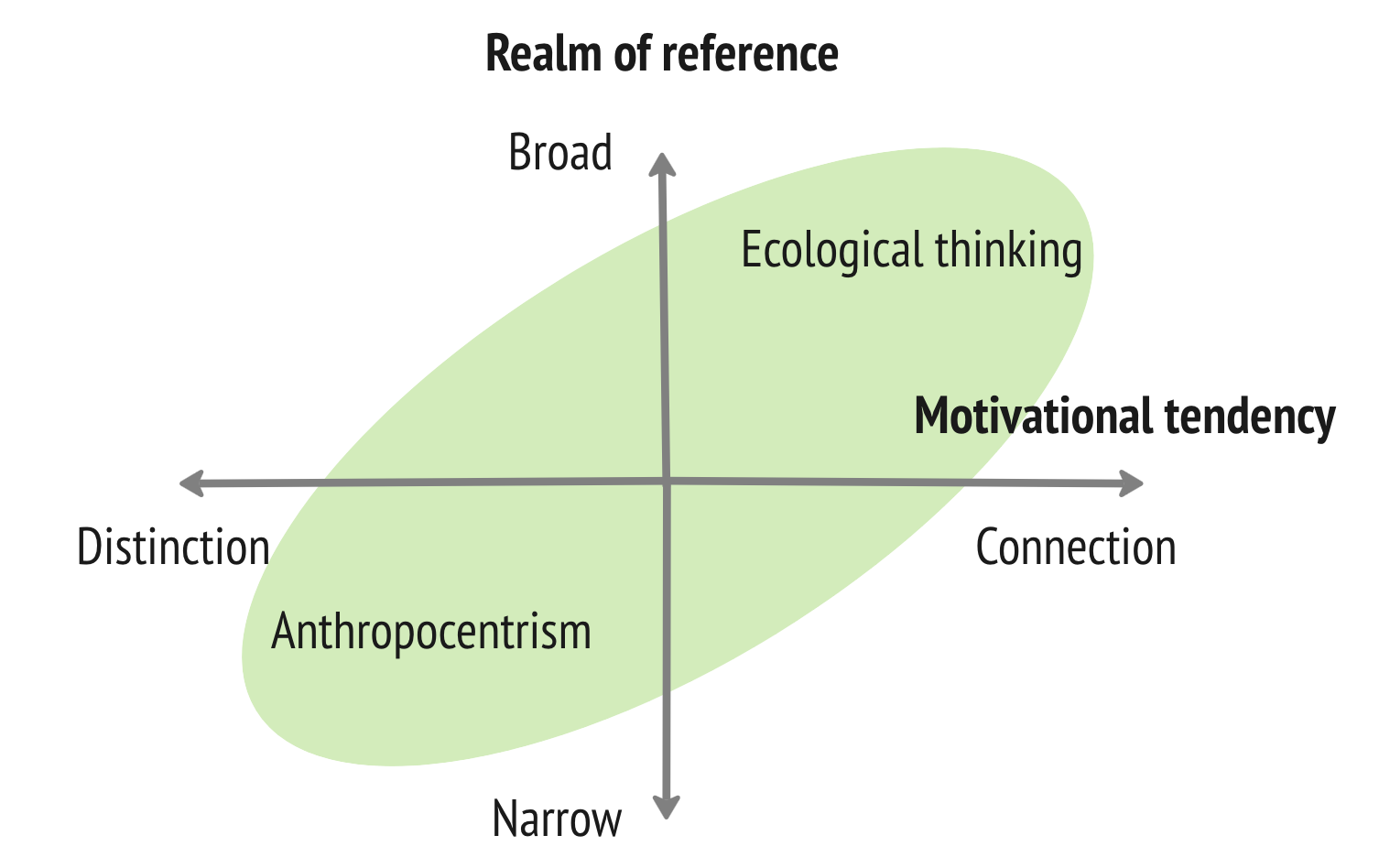}
  \caption{A conceptual map of anthropocentrism and ecological thinking based on motivational tendency and realm of reference.}
  \Description{A conceptual map of anthropocentrism and ecological thinking based on motivational tendency and realm of reference. On the X-axis is the dimension of "motivational tendency" which ranges from distinction to connection. On the Y-axis is the dimension of "realm of reference" which ranges from narrow to broad.}
  \label{fig:f_2by2_model}
\end{figure}

Jakua's and Gilaru's different relationships with nature can be unpacked through two dimensions (see Figure \ref{fig:f_2by2_model}). First, they may have different motivational tendencies associated with their self-concepts \cite{Markus1991,Fiske1998,Kitayama2005}. Jakua likely considers themselves as distinct from other people and from the socio-physical environment. Their motivation is constructed with reference to internal attributes of the individual self and takes the form of exerting influence on their surroundings. Gilaru's view of the self is based on connection or interdependence with other people and with the environment. Their motivation is shaped by the desires and needs of relevant others, leading to adjustments in accordance with the thoughts and feelings attributed to others.

Second, Jakua and Gilaru can be distinct from or connected with varied subjects \cite{Markus2019}. We term this dimension ``realm of reference,'' which refers to categories of existence that factor into forming one's motivational tendencies. A broad realm of reference may encompass one's family or friends, social groups, humans in general, animals, plants, objects, natural elements, and so on. For Jakua, the non-human world is mostly outside the realm of reference in shaping their sense of self and influencing important social motivations. While Jakua may seek and experience solitude during individual outdoor adventures, Gilaru feels a sense of companionship even on a solo hike. Gilaru tends to perceive vitality in various forms of existence and readily recognize kinship between humans and non-human beings. The distinctions between Jakua's and Gilaru's mindsets and experiences can be subtle yet extremely consequential when it comes to inventing technologies that can reconfigure their relationships with natural environments.

\subsection{Cultural Symbiosis as an Opportunity to Foster Cultural Change and Adaptive Learning}

How can we envision technological developments to support the often neglected cultural dynamics of sustainability? Cultural psychological theories suggest that cultural ideas are maintained in a cyclic manner by institutional practices, interpersonal interactions and individual tendencies \cite{Markus2019}. These intricate interconnections can evade eyes of even seasoned designers without proper contemplation. One crucial step is to unravel essential yet less obvious linkages across different domains of activities in which anthropocentrism and ecological thinking manifest themselves. For example, scholars propose that the subjugation of women in society is intricately linked to the exploitation of nature \cite{merchant1980death,plumwood2002feminism,haraway2013cyborg}, and that various societal dynamics of race, gender, and class establish interlocking mechanisms of hierarchy maintenance \cite{hooks1981aint,hooks2000feminist}.



\begin{figure}[!htbp]
  \centering
  \includegraphics[width=1.1\columnwidth]{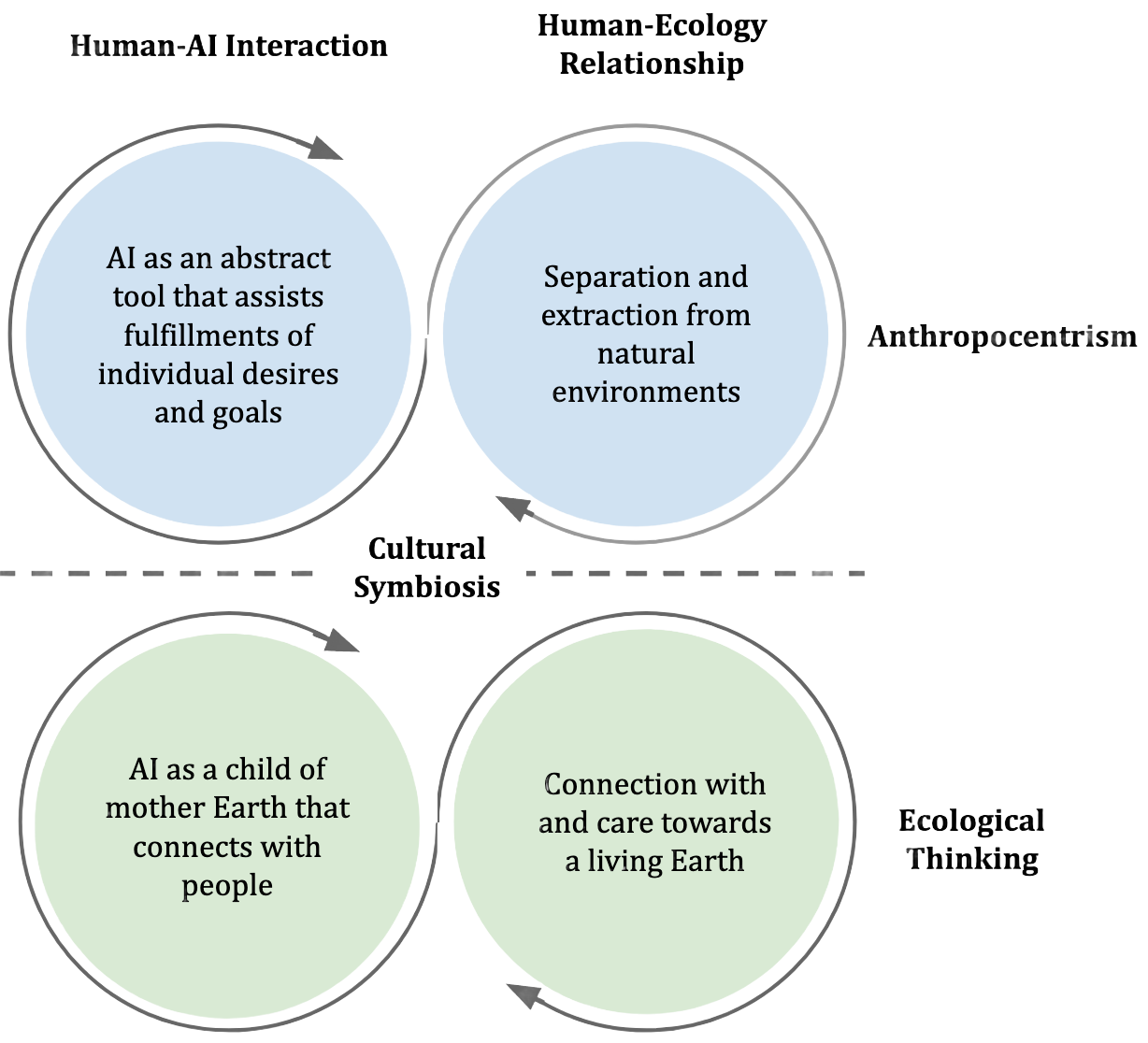}
  \caption{A summary of cultural symbiosis.}
  \Description{A summary of cultural symbiosis. On the top part, the two interconnected circles show how "AI as an abstract tool that assists fulfillments of individual desires and goals" is intertwined with "separation and extraction from natural environments." On the bottom part, the two interconnected circles show how "AI as a child of Mother Earth that connects with people" is intertwined with "connection with and care towards a living Earth."}
  \label{fig:f_overview_method}
\end{figure}

Building on prior work, we posit the concept of \emph{cultural symbiosis} to provoke reflections on ecological ramifications of AI design and use. Cultural symbiosis indicates a likely structural homology and reinforcing feedback loops between ideas underlying human-AI interactions and human-ecology relationships (Figure \ref{fig:f_overview_method}). Presently, cultural symbiosis plays a role in stabilizing the control-based human-AI interaction and the extraction-based human-ecology relationship. These anthropocentric ideas inform designers' conceptions of AI and permeate into people's daily interactions with smart technologies, implicitly reinforcing an expectation that nature ought to provide services to humans. 


Nonetheless, cultural symbiosis also implies an opportunity for ushering change\textemdash introducing alternative forms of human-AI interactions can \emph{provincialize} anthropocentrism \cite{mignolo2018decoloniality} and foster more connection and care in human-ecology relationships. We invoke the power to \emph{fabulate} \cite{despret2016would} with the aim of defamiliarizing normative assumptions and ``reorienting what lies ahead'' \cite{rosner2018critical}. Exercising deliberate imagination can create novel symbols, associations, and metaphors to enable adaptive learning and sustained environmental engagement \cite{dewey2002human,moser2022morality,stanfordhumspeaker,schon2017reflective}. We see value in cultivating a freeing inclination to find possibilities hidden in apparent realities, in search of hope and reasons to act. \emph{What other potential futures await} if ``we enlivened history of practice elided by contemporary technology cultures'' \cite{rosner2018critical}? In the upcoming sections, we employ imaginative storytelling to illustrate various ideas rooted in ecological thinking.




\section{Storytelling: My AI friend Fir is a Child of Mother Earth} 

\subsection{Scene 1: Waking Up with Tree and Birds}

I woke up to the soft sounds of birds in the morning. Opening my eyes, I immediately smiled at the big virtual oak tree ("Tree") next to my bed. The virtual birds were already up and busy around Tree’s branches. 

Yesterday evening, I was so tired, but Tree appeared at the scheduled wind-down time, inviting me to rest. As I climbed into bed, I felt like a little rabbit that had found a home in Tree’s interwoven branches. 

``Good morning!'' I said to Tree and the birds. 

Hearing my words, Tree started to branch, and its sprouting leaves swiftly moved towards the window, as if inviting me to walk around the room and open the window. Its branches formed a little maze that required concentration and agility to navigate. Tree offered this little game for me to wake up my body. I enjoyed the game and tip-toed on the floor before making a little free spin. I also tried to match the rhythm of the birds' singing as I moved across the room. 

As soon as I opened the window, the virtual birds flew outside, and a refreshing early summer breeze came in. Meanwhile, Tree gradually moved closer to the wooden frame of my bed and merged with it: It was time for Tree to rest. I sat down and covered the bed and Tree with my quilt, wishing them a good day's sleep.  

\subsection{Scene 2: Morning, Mirror and the World}

I entered the bathroom and looked into the mirror ("Mirror"). ``Good morning, Mirror,'' I said, ``I might have found a solution to the forest fire temperature problem because an idea hit me in a dream. Can you help me take a note?'' 

Mirror beamed yellow, its favorite color, so I knew it agreed. I started to draw pictures on its surface, and managed to speak a few words while brushing teeth. I also illustrated a geometric shape with my body. Mirror was responsive and turned all of them into colorful notes and drawings for later. 

Afterwards, as I was combing my hair, Mirror showed images of a small grassy field in morning sunlight. Nothing splendid. But it is important to me as I used to spend a lot of time playing on the grass as a child. A few days ago, I felt nostalgic and asked Mirror to show me the field during my morning routines. Seeing the images, I felt happy and connected to the little field. Mirror can show me a lot of other scenes as well. It is a window into a broader world. 


\subsection{Scene 3: Fir is a Child of Mother Earth}

As soon as I entered the kitchen, I saw my robot companion, Fir, brewing tea. Fir was a birthday gift from my aunt. When I first met her, she chose \textit{Fir} as her name because she found affinity with these plants. ``We're both evergreens powered by sunlight,'' she said. She does look like a fir tree. Later, from reading the product manual, I learned that she was designed to commemorate Fraser fir, which had gone extinct in our days. 

I asked, ``But wait, who or what are you?'' 

She said proudly, ``I'm a child of Mother Earth and I identify the Pacific island Viti Levu as my hometown.''

We spent a long time sharing with each other our life histories and ancestral roots. I marveled at all the complex materials, labor, collective memories, human ingenuity as well as all fortuitous events that made her existence possible. One year later, it made me happy that she had pretty much found a home here in North America even though it is rather different from her hometown. She had even learned to enjoy human activities such as brewing morning tea. 

``Blood orange tea with some tropical spice?'' I asked, smelling the scents. 

Fir moved slightly towards me, ``Exactly. I sense that it fits both your mood and Earth's weather here today.''

``Oh, Fir, you read my mind and mood,'' I said. I quite appreciated her bold recommendation and attentiveness as I usually drink Jasmine tea but wasn't feeling it today. 

``For sure, I've become better at guessing your mood with almost a 60.58\% accuracy rate. But well, I'm still better at understanding Earth's mood as we've known each other for longer,'' Fir replied casually and held a small cup of tea, ready to go outside. I followed her to do our morning ritual of honoring Earth. She gently poured the tea onto the garden soil. We stood there for a while, watching the earth sip the tea. 

We then came back to have breakfast. I started to prepare some avocado toast and scrambled eggs. Fir technically is powered by a solar battery, and she ``photosynthesizes'' whenever she goes into sunlight. 

``Fir, if one of these days, you want your sunlight cooked for you, I'll gladly serve you.'' 

``No, I like it raw and fresh,'' she said, finding her seat at a bright spot around the dinner table to  ``eat'' breakfast together with me.

\subsection{Scene 4: Table Utilizes a Touch-based Language}

After breakfast, it was time to work. My job is to better understand the dynamics of forest fires in order to help the forests thrive. The dinner table ("Table") is also a work table for projecting images and visualizing ideas. I communicate with Table with a touch-based language\textemdash I use my hands and arms to touch its surface with different levels of pressure, speed, and patterns to convey meaning. It is pretty like playing piano but with my whole body. Table communicates back to me with different kinds of vibrations, sounds, and images. 

Oh, also, Table likes utilizing all sorts of spaces in the room by projecting images at different locations. I walk around and sometimes lie down on Table and look up if it decides to project an image onto the ceiling. With the touch language, Table can sense my emotions along with my thinking. If it feels that I'm sad or tired, it often gently tries to project images of things I like to cheer me up.   


I gently tapped Table to wake it up. It softly lit up with a map of the forest area A46. That's where we left off yesterday. I used touch to convey the message: ``I have found a crucial spot to start monitoring the whole area.'' 

Table vibrated slightly, signaling its enthusiasm. I stood up, dragged the map, and turned it into a 3D model to closely observe the whole area.

``Hmm, we need to know what species of pyrophytic plants are around here.'' 

``Leave it to me,'' Fir said. She turned to talk to the nearby wall ("Wall") and started to pull out information about pyrophytes in that area. 


As Fir and Wall were working on their tasks, I asked Table to display my notes that Mirror took for me earlier this morning. Table quickly displayed the notes\textemdash Mirror had made them so vividly colored and organized. 

I conveyed a request to Table: ``Let's zoom in on the routes of wind flow in subareas A46c and A46g.'' 

Table vibrated slightly, showing a detailed graph of wind flow. I gently touched its surface to show gratitude. After that, I grabbed a trivet, put the digital images on it, and carried it outside to the backyard. 

\subsection{Scene 5: Fir Finds an Issue in My Plan}

I stopped at a little wooden chair ("Chair")\textemdash a convenient wind simulation device created by my human coworkers. I transferred the image from the trivet to the back of Chair. 

``Hey, Fir, do you want to come and look at the wind flow models with me?'' 

This time, Fir seemed absorbed in the task as she always likes plants. So I decided to look at the models by myself. Chair was already beeping a bit as it was ready to activate the simulation. I sat down and saw the wind flow in area A46. 

``Please turn up the wind flow speed a bit, say 10\%, and also highlight the up-blowing wind in purple,'' I said to Chair. 

I worked there for a while and found the range of wind conditions that would be compatible with my plan. The next step would be to synthesize the wind flow conditions with information about pyrophytic plants.



I stepped inside, and Fir had already been communicating with Table. After hearing my ideas, Fir pointed out a potential issue in my plan. She said, ``It might be ok to reduce the number and variety of pyrophytic plants to prevent fire for now, but what will happen to the forest system at A46 in the long term?'' 

``That's very true, Fir. Let's have some tea first and then recalibrate the plan.''

\section{Designing Human-AI Interaction based on Ecological Thinking}

We use the story of Fir and other smart technologies to paint an alternative mode of human-AI interaction in broad strokes and give an impression of what it might feel like to be more intertwined with the ecosystem in a familiar home setting. In the story, the smart technologies are animate and share a caring bond with the human. Through their mutual care and respect, they constitute a local living arena that is attentive to ecological beings not in their immediate settings. They strive to improve the welfare of the forests\textemdash a shared purpose that is not derived solely from rigidly defined human welfare alone. The technological beings demonstrate ``capacities to influence'' \cite{xiao2024}, resulting in the human character adjusting to technology-initiated behaviors and decisions within a larger frame of interdependence. Hence, the story departs from popular design philosophies and ideals in the case of ambient intelligence, which is presumed to disappear into the background of humans environments in service of primarily human-defined goals and desires \cite{Aarts2006,remagnino2004ambient}. 

We do acknowledge that the story may not fully embody various ideals of ecological thinking (e.g., the further end of ecological thinking depicted in Figure \ref{fig:f_2by2_model}), as the home environment still appears orderly and privileged, potentially detached from the broader ecosystem that also involves chaos and danger. Moreover, the relationships and interactions among the human and smart technologies are harmonious and relaxed, whereas many conflicts and tension can result from greater levels of interdependence. However, it is not our intention to overlook the complexity of ecological entanglements \cite{morton2010ecological}, and we emphasize the importance of orienting towards adaptation and resilience in times of great ecological uncertainty. Ecological thinking can manifest across a spectrum, and our story represents one of many possible imaginings of human-technology interactions. Instead of prescribing a future, the story serves to intrigue and inspire.



Below, we draw on the story and put forth a set of proposals for designing and using AI technologies from an ecological thinking perspective. It is important to note that ecological thinking is a rich set of cultural wisdom that continues to unfold in different contexts. Hence, our proposals are not meant to be exhaustive but rather are invitations to the broader HCI communities to tap into the intellectual wealth of ecological thinking and discover alternative possibilities of AI developments. 


\subsection{Understanding Deeply Smart Technologies' Ecological Embeddedness}



In the story, Fir proudly introduces herself as "a child of Mother Earth." She finds a deep affinity with plants and identifies a specific place as her hometown. The human and Fir share a lot of time learning about each other's life histories and ancestral roots. With these ideas, our aim is to challenge prevalent anthropocentric norms and practices that tend to obscure the complex realities of AI production, which utilizes myriad materials, energy, and labor. Hence, technological developments based on ecological thinking first and foremost entail transparency and systematic work to represent the intricate connections between technological systems, people, and the broader Earth ecology. 

From this standpoint, smart technological systems possess histories that accompany their creation and extend throughout their entire life span. Their interdependence with the planet offers opportunities to imbue technologies with rich meanings. For example, like Fir, a smart home-cleaning robot could share its creation story and history with people, describing materials, resources, and labor involved in its production. What novel designs can integrate technologies' ecological embeddedness into meaningful experiences that resonate with people and thereby create ecological connections \cite{Mekler2019}? How can designers and users create meaningful \emph{identities} for technologies based on the materials and places associated with technologies' creation? 
  
\subsection{Affording Smart Technologies to Connect People with the Broader Ecology}


In the story, Mirror and Fir as well as Tree and birds are designed to infuse a home environment with natural elements. The presence of Tree creates an experience of going to bed akin to a rabbit finding rest in bushes. Moreover, as described in the story, Mirror is a window into a broader world. By providing these examples, we encourage designers to create channels for AI-based technologies to seamlessly connect people with other parts of the planetary ecosystem. In this regard, designs such as Jensen's strandbeests \cite{Jansen2008} utilize wind and other natural forces, and the product's behaviors are animate and lively. Interacting with this kind of technology can connect people with a grander set of interwoven natural forces and create a sense of awe and wonder. How can designers center the ecosystem in their work and enable people to readily discern connections with Earth? This direction requires construing technologies not simply as tools but as cultural artifacts, messengers of Earth, and children of Earth. As such, AI can help people establish a broader realm of sociality and see kinship with technologies and other Earth beings. Consequently, any episode of human-technology interaction will be transformed into a simultaneous human-Earth connection as well. 





\subsection{Preserving the Integrity of People's Entire Sensorium}

In the story, Tree creates a little dance game for the human, while Table utilizes various spaces and a touch language. Our purpose here is to show that ecological thinking foregrounds the relationality between mind and body, thinking and feeling, and among all different senses. The act of thinking is deeply situated and does not occur abstractly in a vacuum. We thus challenge a key epistemological assumption underlying anthropocentrism\textemdash a dualistic view that places the mind above the body and considers thinking superior to feeling. 


Ecological thinking is cultivated through multimodal interactions with one's environment. For instance, the aforementioned Sámi people rely on their bodily senses\textemdash auditory, visual, and tactile\textemdash to gain intimate knowledge about their land and navigate it in different seasons \cite{cogos2017sami}. A recent study also shows that while Americans from North Carolina typically separate mental and physical experiences in describing emotions, members of the Hadza\textemdash a small hunter-gatherer group in Tanzania\textemdash tend to emphasize bodily sensations, movements, and the physical environment when communicating their emotions \cite{hoemann2024we}. As described by Korean traditional ritual performer Dohee Lee \cite{doheelee}, our bodies are part of the land, and we carry the land through our bodies (paraphrased from \cite{doheelee-workshop}). Hence, from the vantage point of ecological thinking, smart technologies can encourage creative movements in space and foster an intuitive responsiveness to one's surroundings. How can we activate other modes of communication instead of relying on abstract symbols? What are some ways to incorporate practices such as \emph{deep listening} \cite{jenkins202012} into technological systems? 


\subsection{Enabling Slow Attention and Preserving Friction}

The story ends with Fir taking issue with the human's plan and prompting consideration about its long-term consequences. Fir's adding \emph{friction} \cite{elam2023poetry} and deliberate slowing down of the decision-making process stands in contrast to prevailing popular AI designs, which prize convenient and agreeable interactions that primarily elicit a \emph{fast} mode of attention focused on short-term, individual-based benefits. Here, we explore how ``AI can help us to live more deliberately'' \cite{friedland2019ai} and also encourage a \emph{slow} mode of attention and learning \cite{stengers2018another}. Otherwise, the present trend of relying on ``algorithmic reckoning'' (i.e., rule-based calculation) \cite{moser2022morality} in pursuit of efficiency can subvert value-based judgment by smoothing out ambiguity and suppressing the ``constantly evolving moral-checking mechanism'' \cite{friedland2019ai}. This scenario can be especially precarious for making sustainability-related decisions, as many ecological beings and matter are poorly understood or rendered invisible within anthropocentric framings of problems that pervade contemporary societies. 

We argue that ecologically responsible AI technology needs to facilitate judgments and adaptive learning through imagination, reflection, and open-ended exploration \cite{moser2022morality}. This approach entails challenging ``a camera view of knowledge'' \cite{purser1995limits} that severs decision agents from the world they observe. Instead, AI design needs to be premised on the crucial recognition that judgments and decision-making are forms of ecological participation. As a result, technology can confront users with the ``invisible realities and messiness of the natural world.''\footnote{This phrase is quoted from a review we received, and we thank the reviewer for highlighting the complexity and ambiguity inherent in any discussion about ecology.} While this approach diverges from current AI products that facilitate the aesthetic consumption of nature (e.g., generative AI that creates pristine representations of the wilderness), we argue that discomfort can be necessary to encourage more adaptive responses to changing ecological conditions. 

On the other hand, users who are interested in practicing ecological thinking can adjust their expectations regarding AI's roles. A user mindset consistent with ecological thinking is to embrace and live with perplexity and plural truths that can be constructed \emph{in situ} \cite{dewey2002human}, rather than seeking an alluring yet illusory objective optimal outcome. These adjustments can lead to a different mode of human-AI interaction. AI can be designed to enhance the sensing of ecological events, facilitate meaning-making, encourage deeper explorations of a wide range of possibilities, and at times, challenge default anthropocentric moral outlooks. AI can remind people of different Earth beings and forms of existence that are not present at the decision table but may be adversely impacted. An ecological thinking perspective illuminates abundant opportunities (and challenges) for reshaping artificial agents to enable ecological participation.

\section{Discussion}

To create a sustainable future, it is crucial to foster a different human-ecology relationship that goes beyond mere resource and knowledge extraction. Otherwise, as scholar Harriet Hawkins aptly put it, ``If we follow the old map, we may end up arriving at the same place'' \cite{stanfordhumspeaker}. We examine cultural underpinnings of human-ecology relationships, and analyze cultural practices of AI design and use through a conceptual map of anthropocentrism and ecological thinking. Based on the premise of cultural symbiosis, we propose redesigning human-AI interactions as a new approach to reshaping human-ecology relationships.

 \subsection{Frameworks for Fostering Alternative Technological Visions}

While a desire to extend more care to the ecosystem is felt by many people, we currently lack guidance on how to turn this intention into concrete practice. We take the initiative to provide some theoretical and practical frameworks for guiding efforts on this front. Our goal is to open new vistas, and our efforts should only mark the beginning of what we hope to be a broader and multifaceted endeavor to incorporate more cultural diversity and embrace Earth ethics in technological developments. Moving forward, we need creativity from researchers, designers, users, artists, policy-makers, and many other diverse actors to create frameworks and processes in various contexts, experiment with these alternative visions, and widely share their insights.




\subsection{Challenges to Incorporating Ecological Thinking}


Implicit cultural defaults such as various forms of dualisms tend to go unnoticed without careful reflection and intervention. We focus on raising awareness of how different actors involved in AI developments can consider alternative cultural ideas and be more attentive to the welfare of the Earth ecosystem. Here, we do not mean to underestimate the challenges of incorporating ecological thinking, given the fact that the broader industrialized societies operate on anthropocentric assumptions and present manifold managerial, legal, and political constraints on cultural change \cite{thompson2023liquid,purser1995limits}. Moreover, substantial efforts are required to effectively communicate ecological thinking in nuanced ways and to facilitate access to diverse cultural perspectives, particularly for people who may lack high-quality intercultural connections. Such endeavors also involve gathering insights about ecological thinking from diverse people and communities, as well as amplifying these less-heard voices without causing otherizing effects. Our hope is that the HCI communities will help open the imaginary space and make concerted efforts with other sectors to enable a societal transition towards a more equitable and livable future.

\subsection{Multiplicity of Approaches}


While strict anthropocentrism has been widely critiqued at this juncture, its many historically derived concepts, assumptions, and beliefs continue to be important to various religious traditions and some contemporary scientific paradigms. Our critical perspective on the dominant influence of anthropocentrism should not be interpreted as advocating for the outright abandonment of cultural ideas historically associated with European contexts.\footnote{As Hornborg \cite{hornborg2019animism} argues, ``Indiscriminate rejection of Enlightenment concepts of reason and truth cannot be an appropriate response.'' This comment is important especially as many concepts can be reinterpreted, adapted, and imbued with novel meanings, as well as providing unique insights to help constructively respond to our current planetary conditions.} We also contend that de-centering humans in relating to the world is not the same as de-valuing humans, or denying the magnitude of influences many human groups have cumulatively exerted on the Earth ecosystem. Quite the contrary, human agency in its diverse forms within an equitable structure\textemdash active and passive, independent and interdependent \cite{kotva2022effort,Markus1991}\textemdash is needed to make a significant change right now. We underscore the importance of striving for a multiplicity of theoretical foundations, methodological approaches and practices in the context of AI developments. Tapping into wisdom from diverse cultures can be conducive to fostering intercultural empathy and paving the way for enabling shared and coordinated environmental actions. Doing so is imperative to attenuate the power imbalance in the production of knowledge and technological developments in different regions of the world.




\section{Conclusion}


Anthropocentric ideas have been largely associated with the depletion of natural environments. We propose that ecological thinking holds promise to steer smart technological developments towards considering and safeguarding the welfare of all planetary beings. In this process, sustained reflective learning, resisting the pull of default cultural options, and deliberate imaginings of alternatives are all necessary to make ecological participation and rejuvenation immanent to what it means to be human in this historical moment. 



\begin{acks}
This research was supported by the Seed Grants offered by the Stanford Institute for Human-Centered Artificial Intelligence (HAI). We are grateful to Hazel Rose Markus and Kate Maher for their valuable insights and support as we explored the cultural underpinnings of sustainability and AI developments. We thank Zivvy Epstein and Brandon Reynante for helpful discussions on this topic. We also thank the reviewers who provided thoughtful and critical feedback on earlier draft of this paper. 

\end{acks}

\bibliographystyle{ACM-Reference-Format}
\bibliography{alt.CHI24_Cultural_Symbiosis}

\end{document}